\documentclass[english,showpacs,prl,twocolumn]{revtex4}
\pdfoutput=1
\usepackage[T1]{fontenc}
\usepackage[latin9]{inputenc}
\usepackage{babel}
\usepackage{natbib}
\usepackage{amsmath}
\usepackage{graphicx}
\usepackage[unicode=true,pdfusetitle,
 bookmarks=true,bookmarksnumbered=false,bookmarksopen=false,
 breaklinks=false,pdfborder={0 0 1},backref=false,colorlinks=false]
 {hyperref}

\makeatletter
\@ifundefined{textcolor}{}
{%
 \definecolor{BLACK}{gray}{0}
 \definecolor{WHITE}{gray}{1}
 \definecolor{RED}{rgb}{1,0,0}
 \definecolor{GREEN}{rgb}{0,1,0}
 \definecolor{BLUE}{rgb}{0,0,1}
 \definecolor{CYAN}{cmyk}{1,0,0,0}
 \definecolor{MAGENTA}{cmyk}{0,1,0,0}
 \definecolor{YELLOW}{cmyk}{0,0,1,0}
}

\makeatother

\begin{document}

\title{Entropic commensurate-incommensurate transition }

\author{Nikolai Nikola, Daniel Hexner, and Dov Levine}

\affiliation{Department of Physics, Technion-IIT, 32000 Haifa, Israel}
\begin{abstract}
The equilibrium properties of a minimal tiling model
are investigated. The model has extensive ground state entropy, with
each ground state having a quasiperiodic sequence of rows. It is found
that the transition from the ground state to the high
temperature disordered phase proceeds through a sequence of periodic
arrangements of rows, in analogy with the commensurate-incommensurate
transition.  We show that the effective free energy of the model resembles the Frenkel-Kontorova 
Hamiltonian, but with temperature playing the role of the strength of the substrate
potential, and with the competing lengths not explicitly present in the basic interactions.
\end{abstract}


\maketitle
Tilings provide a simple means to model systems with both simple and
complex ground states. As statistical mechanical models, they include
such systems as the Ising and Potts models, as well as other models
with discrete spin variables\cite{fn1}. The class of tilings, however,
is much larger\cite{tilings_and_patterns}, and includes the various
non-periodic Wang tilings\cite{fn2}, the quasiperiodic Penrose tilings,
the asymptotically isotropic Pinwheel tiling\cite{pinwheel}, and
the recently discovered generalizations of the Rudin-Shapiro sequence,
which are neither periodic nor quasiperiodic\cite{Wolff}. 

The statistical mechanical behavior of tiling models is rich, and,
as yet, largely uncharted. Apart from results which may be transcribed
from discrete spin models, only a very small number of systems have
been studied. First, a model based on the Amman set of 16 Wang tiles
with quasiperiodic ground states was studied by several authors \cite{leuzzi_thermodynamics_2000,rotman_finite-temperature_2011,koch_modelling_2010}.
It appears that this model undergoes a continuous phase transition
from a disordered state to a quasiperiodic phase, and its non-equilibrium
behavior was studied in the context of spin-glasses. A variation of
the model allowing more complicated interactions and vacancies shows
a first order transition\cite{aristoff_first_2011}. Hierarchical
tilings\cite{miekisz_microscopic_1990} have also been studied, and
a very recent model\cite{byington_hierarchical_2012} possesses limit-periodic
ground states which undergo a series of phase transitions where motifs
of ever larger scales order as the temperature is lowered. Finally,
we note some recent studies \cite{sasa_pure_2012,sasa_statistical_2012}
on models with large number of degenerate disordered ground states
aimed at studying glasses. 

In this paper we study the equilibrium behavior of a model based on
the 13-tile Kari-Culik (KC) set \cite{culik_ii_aperiodic_1996,kari_small_1996}
of Wang tiles, both numerically and analytically. The KC set is the
smallest known \emph{aperiodic} set - it is the smallest set of tiles
which can tile the plane, but not periodically. Allowed juxtapositions
of tiles are enforced by matching rules, and these in turn induce
a Hamiltonian: Every matching rule violation is penalized by a positive
energy, while allowed matchings have zero energy. In what follows,
we shall denote the energy cost of mismatching adjacent vertical or
horizontal edges by $J_{x}$ or $J_{y}$, respectively. 

We shall argue that the equilibrium behavior of this system is analogous to the Frenkel-Kontorova (FK)
model\cite{chaikin_principles_2000} of the commensurate-incommensurate (CI) transition.  The FK model describes a chain of masses which are connected by springs and subjected to a periodic substrate potential. It exhibits rich behavior due to competition between these two interactions, each of which favors ordering with a different wavelength.  However, in marked contrast with the FK model, where the favored length scales are present in the Hamiltonian, in the KC model they emerge spontaneously, from only nearest-neighbor interactions.  Moreover, the role of substrate potential strength in the FK model is played by the temperature in the KC model.
In this sense, the KC system exhibits an {\it entropic CI} transition.

\begin{figure}
\includegraphics[bb=170bp 0bp 612bp 219bp,scale=0.31]{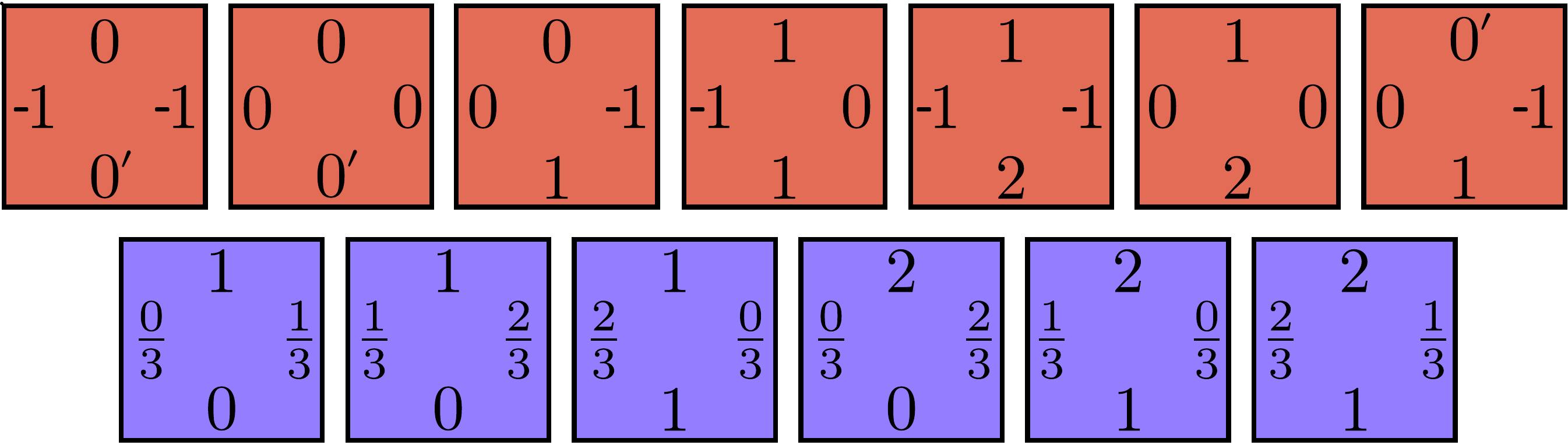}

\caption{The $13$ KC tiles. The seven upper tiles are type \emph{A}, and the
six lower tiles are type \emph{B.} Note that $\frac{0}{3},\,0$ and
$0'$ are considered different markings.}

\label{tiles}
\end{figure}

As indicated in Figure \ref{tiles}, the 13 KC tiles may be divided
into two groups, which we will call types \emph{A} and \emph{B}. The
markings on the edges of the tiles indicate the matching rules - abutting
edges of adjacent tiles have the same markings in a perfect tiling.
It is readily seen that in an undefected tiling, a given row can consist
of \emph{A} or \emph{B} type tiles only, with no mixing, and thus,
we may characterize a row as being \emph{A-type} or \emph{B-type. }

The KC tiling differs from other tilings studied in that it is not
generated by recursive substitution (inflation)\cite{tilings_and_patterns},
and by the fact that it has a ground state degeneracy with extensive
entropy\emph{.} This should be contrasted with tilings created from
a simple inflation rule where the degeneracy scales as a power of
the system size\cite{koch_modelling_2010}. All the ground states,
however, are characterized by a quasiperiodic arrangement of \emph{A-}type
and \emph{B-}type\emph{ }rows. The finite temperature behavior of
this model is striking - as the temperature is increased from zero,
the rows, still identifiable as \emph{A} or \emph{B} type, order periodically
with decreasing period. At high enough temperatures, the rows lose
their \emph{A} or \emph{B} identity, and the system becomes disordered. 

The proof that there exists a zero-energy ground state is equivalent
to showing that a perfect tiling exists. To do this\cite{Eigen},
we note that there are two numbers which characterize the $n^{th}$
row in a perfect tiling - the ``frequency'' $\alpha_{n}$ and the
number $q_{n}$ which are related through the mapping

\begin{equation}
\alpha_{n+1}=q_{n}\,\alpha_{n}\label{eq:mapping}
\end{equation}

where 

\begin{equation}
q_{n}=\begin{cases}
2 & \;\;\frac{1}{3}\leq\,\alpha_{n}<1\\
\frac{1}{3} & \;\;1\leq\,\alpha_{n}<2
\end{cases}\label{eq:q_of_m}
\end{equation}
Note that we shall take \emph{n }to increase in the \emph{-y }direction.

\begin{figure}
\includegraphics[scale=0.25]{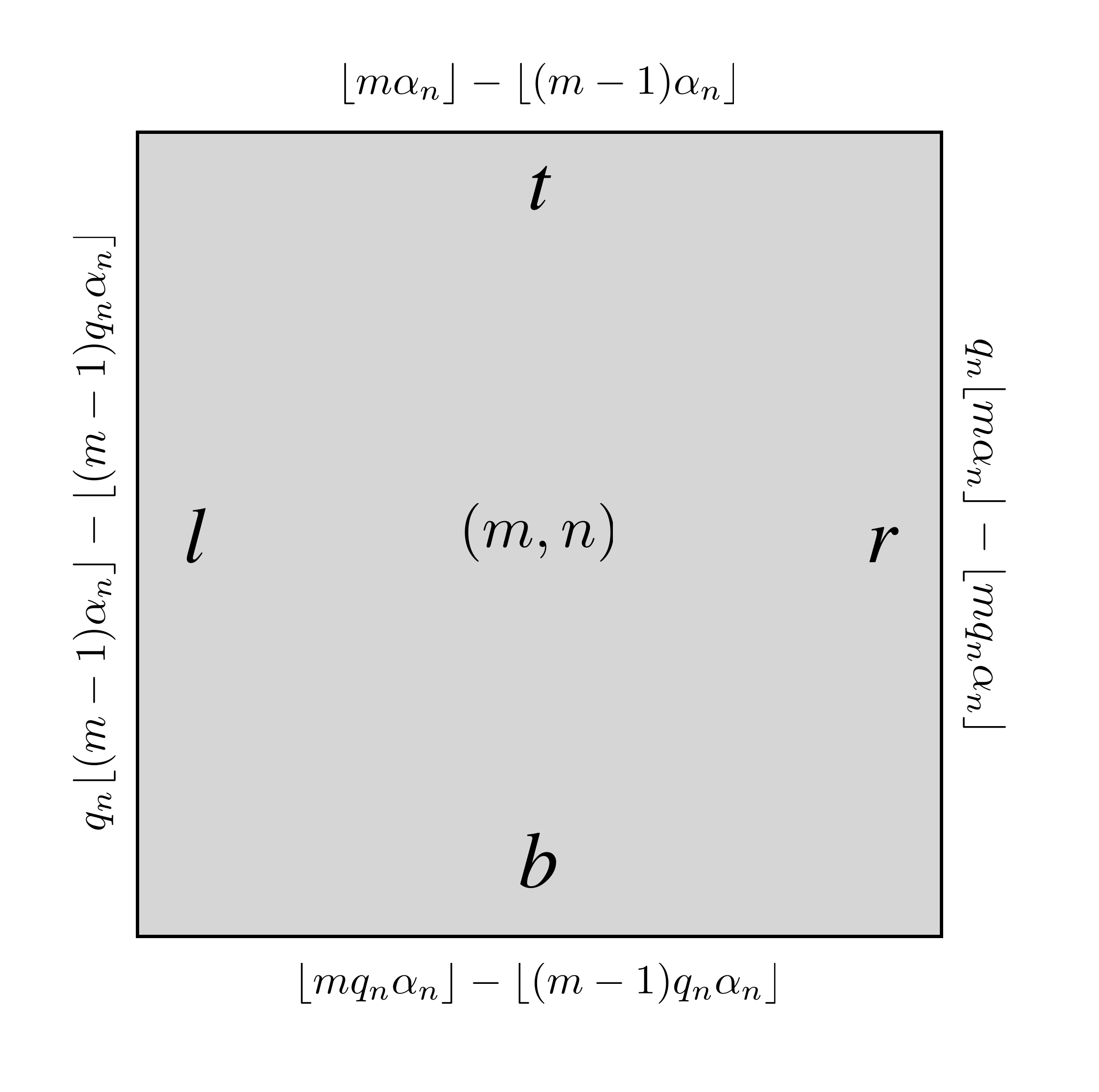}

\caption{The marking scheme for the tile at (\emph{m,n}). Note that \emph{n}
increases in the \emph{-y} direction. }

\label{basic_tile}
\end{figure}

The reason for dividing the tile set into the \emph{A} and \emph{B}
groups becomes apparent if we denote the markings of the \emph{top},
\emph{bottom}, \emph{right}, and \emph{left} edges of a tile by \{\emph{t,b,r,l}\},
as shown in Figure \ref{basic_tile}. When needed, we shall indicate
the position of the tile as a subscript; thus $t_{m,n}$ refers to
the marking of the top edge of the tile centered at (\emph{m,n}),
\emph{etc}. It is easily verified by inspecting Figure \ref{tiles}
that the markings on each tile satisfy the relation

\begin{equation}
q_{n}\, t_{m,n}+l_{m,n}=r_{m,n}+b_{m,n}\label{eq:sym}
\end{equation}
 with a tile of type $A$ or \emph{$B$} having $q_{n}=2\,$ or $\frac{1}{3}$,
respectively. The mapping $\alpha_{n+1}=q_{n\,}\alpha_{n}$ has no
periodic points, since $\alpha_{n+q}/\alpha_{n}=2^{p}/3^{q-p}\ne1$
for any positive integers $p$ and $q$; this implies that the resultant
tilings are not periodic. The $\alpha_{n}$ are distributed densely,
but not uniformly, in the range $\left(\frac{1}{3},2\right)$. 

To show that a perfect tiling exists, we must solve Equations \ref{eq:sym}
for all \emph{m,n }with $q_{n}$ derived from Equations \ref{eq:mapping}
and \ref{eq:q_of_m} whilst demanding that $b_{m,n}=t_{m,n+1}$ and
$l_{m,n}=r_{m-1,n}$. It is easily verified that the markings

\begin{eqnarray}
t_{m,n} & = & \left\lfloor m\alpha_{n}\right\rfloor -\left\lfloor (m-1)\alpha_{n}\right\rfloor \label{eq:t&r}\\
r_{m,n} & = & q_{n}\left\lfloor m\alpha_{n}\right\rfloor -\left\lfloor mq_{n}\alpha_{n}\right\rfloor \nonumber 
\end{eqnarray}
constitutes a solution, where $\left\lfloor x\right\rfloor $ is the
greatest integer less than or equal to $x$ (see Figure \ref{basic_tile});
this is one example of a perfect tiling.

To facilitate later discussion, we shall employ a useful mapping.
Let us define the variable $\phi_{n}=n\omega_{0}$ , where $\omega_{0}=log2/log\left(6\right)\approx0.3868$.
Decomposing $\phi_{n}$ into its integer and fractional parts gives 

\begin{equation}
\phi_{n}=\lfloor n\omega_{0}\rfloor+\left\{ \phi_{n}\right\} \label{eq:Phi}
\end{equation}
where $\left\{ \phi_{n}\right\} $, the fractional part of $\phi_{n}$,
is defined through its relation to $\alpha_{n}$ by
\begin{equation}
\left\{ \phi_{n}\right\} =\frac{log\left(\alpha_{n}\right)+log\left(3\right)}{log\left(2\right)+log\left(3\right)}\label{eq:theta}
\end{equation}
This maps $\alpha_{n}\in[\frac{1}{3},2)$ onto $\left\{ \phi_{n}\right\} \in[0,1)$,
and allows us to obtain\cite{futurepub} an explicit formula for $q_{n}$
: 
\[
q_{n}=\left(\frac{1}{3}-2\right)\left(\left\lfloor \left(n+1\right)\omega_{0}\right\rfloor -\left\lfloor n\omega_{0}\right\rfloor \right)+2
\]
Sequences of this type are called \emph{Sturmian sequence}s, and are
well known in the context of automatic sequences\cite{allouche_automatic_2003}.
Here $\omega_{0}$ is irrational, and this gives a quasiperiodic sequence
of the two ``letters'' $2$ and $\frac{1}{3}$, with the consequence
that in the ground state the rows appear in a quasiperiodic sequence.

\begin{figure}
\includegraphics[scale=0.38]{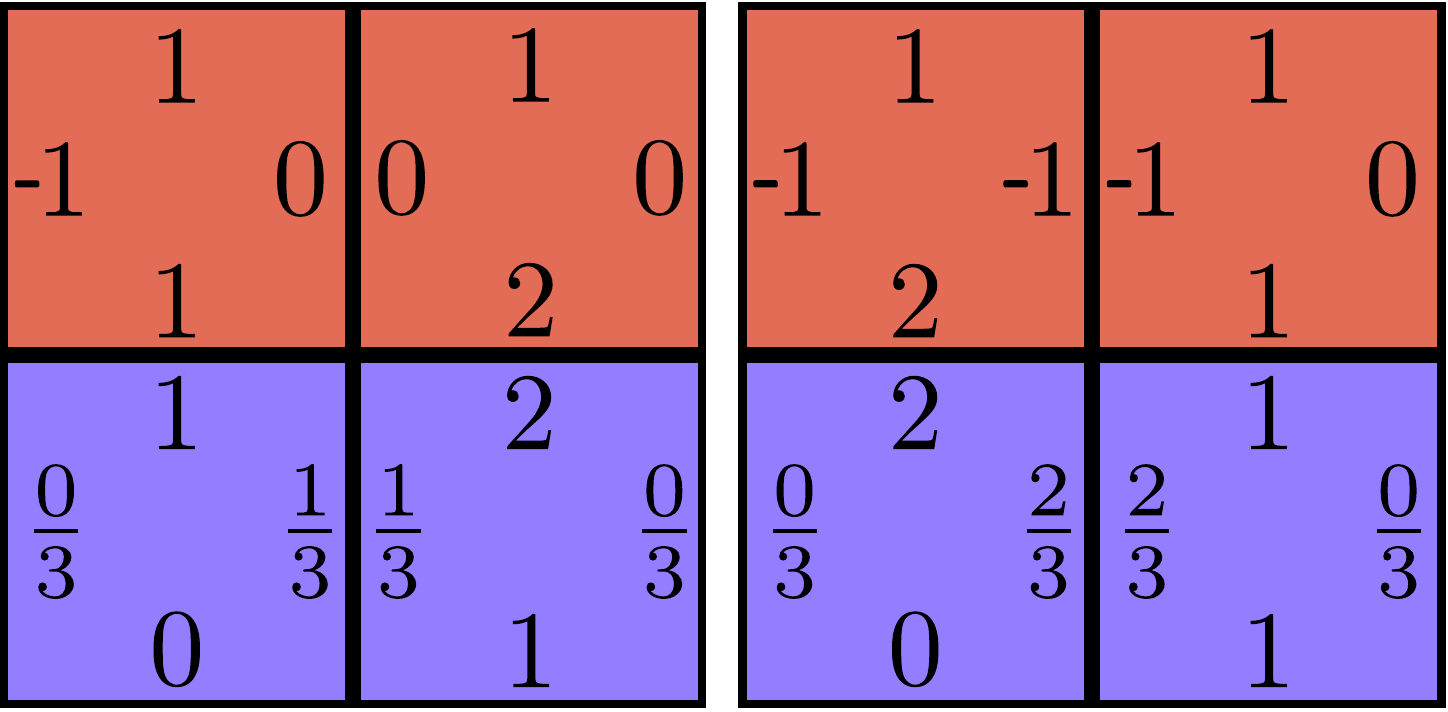}

\caption{An example of two different 2x2 patches with the same outer markings;
these may be exchanged with no matching rule violations.}

\label{swaps}
\end{figure}

As noted above, this ground state is only one of many, and in fact,
there is an extensive ground state entropy\cite{futurepub}. This
may be inferred from Figure \ref{swaps}, where two patches with the
same markings on their exterior are presented (there exist larger
patches with the same property as well). This means that starting
from some given ground state, we may obtain another by randomly exchanging
the patches shown in Figure \ref{swaps} (and any other pair of patches
with the same exterior markings), provided, as we have verified\cite{futurepub},
that they appear with a finite density in the ground state. This implies
that almost all of the ground states are disordered in the sense that
their patch entropy\cite{kurchan_levine} scales as the patch size
for large enough patches. This notwithstanding, in all the ground
states, the rows are arranged in a quasiperiodic sequence of \emph{A-type
}and \emph{B-type,} and it is also true that $\alpha_{n}$ is unchanged
for each row\cite{futurepub}, which is relevant to what follows.

\begin{figure}
\includegraphics[scale=0.38]{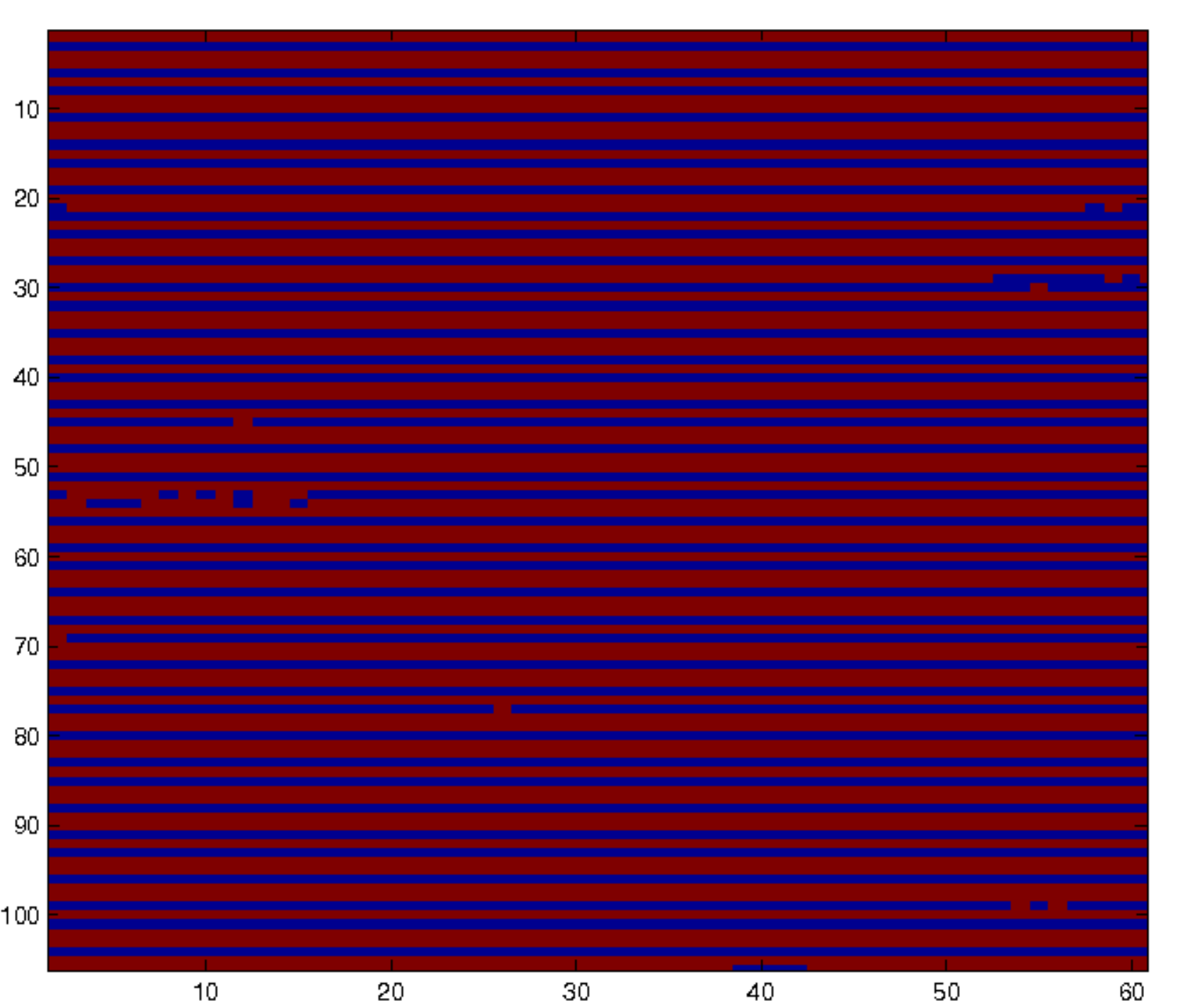}

\caption{Finite temperature configuration averaged over time exhibiting 3/8
periodicity. The row structure is evident, with colors as in Figure
\ref{tiles}. Here $L=150$, $J_{x}=J_{y}$, and $T=0.304$.}

\label{period8}
\end{figure}

Numerical studies of this model show that as the temperature is raised
from 0, it goes through a series of phase transitions, where the \emph{A-B}
sequence of rows is periodic, and where the period decreases with
increasing temperature. In analogy to the CI transition, we shall
refer to these periodic phases as \emph{commensurate phases. }In Figure
\ref{period8}, we show a portion of a time averaged configuration
from a $150\times150$ system (with $J_{x}=J_{y}$) at $T=0.304$
(in units of $J_{x}$), which was obtained by parallel tempering,
where the color coding is as in Figure \ref{tiles}. The \emph{A-B}
sequence is periodic with period 8, with 3 \emph{B} rows per period.
Characteristic defects are also present. At high enough temperature,
of course, the rows lose their \emph{A} or \emph{B} character, and
the system goes to a disordered phase. 
 
\begin{figure}
\includegraphics[scale=0.35]{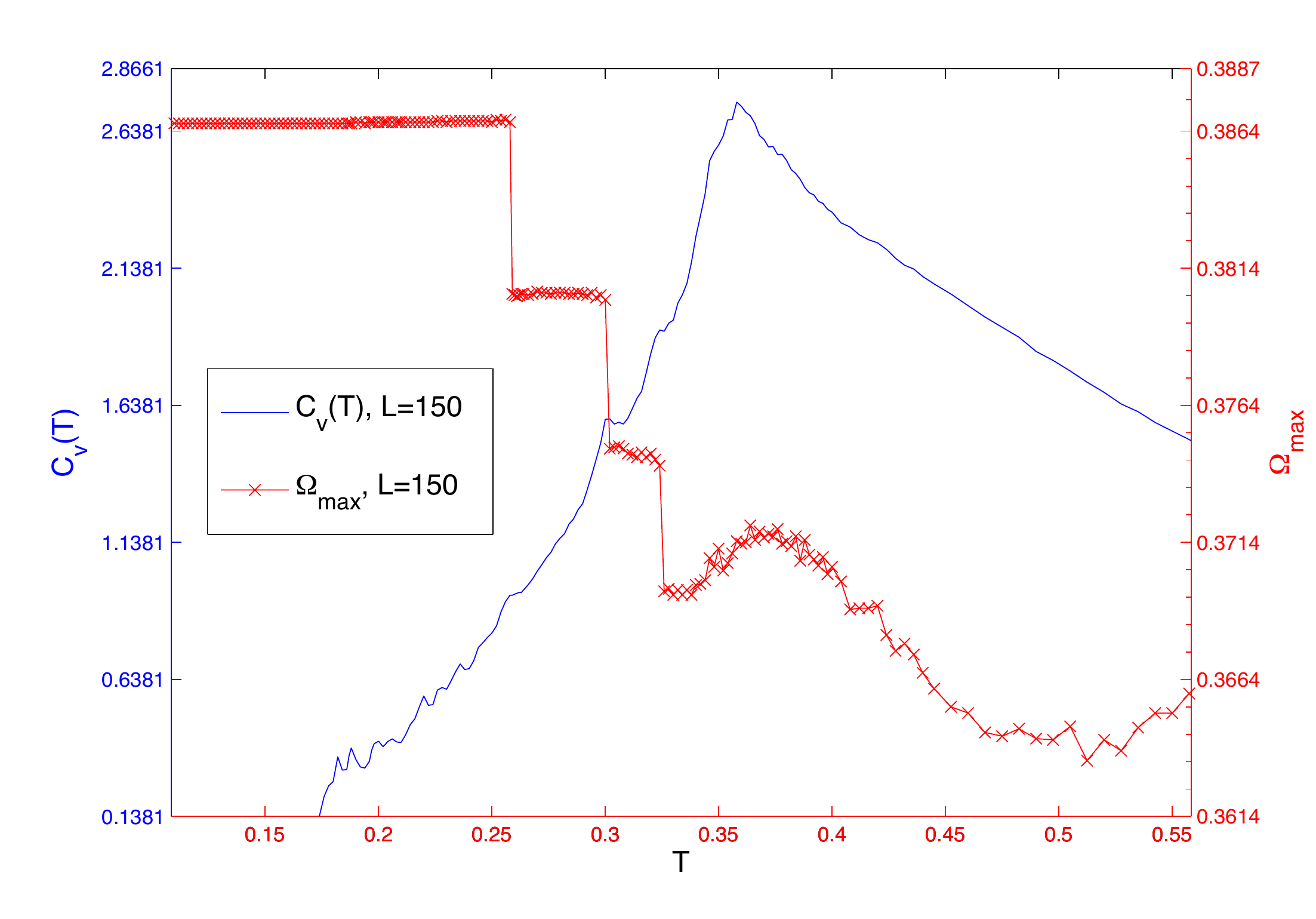}

\caption{$\Omega_{max}$ and $C_{V}$ as a function of temperature, for a $150\times150$
system with $J_{x}=J_{y}$. \textbf{.} }

\label{omega_max}
\end{figure}

The different phase transitions are best traced by the winding number
$\Omega=\frac{N_{B}}{N_{A}+N_{B}}$. Since at low \emph{T }the rows
are essentially pure type \emph{A }or \emph{B}, $\Omega$\emph{ }essentially
counts the fraction of type \emph{B }rows in the system.\emph{ }For
an ensemble of systems at any given temperature, $\Omega$ may take
a variety of values, where the distribution has a well-defined maximum
value, $\Omega_{max}$, which is typically close but not identical
to the average $\langle\Omega\rangle$ %
\footnote{The distinction between $\Omega_{max}$and $\langle\Omega\rangle$
will vanish in the thermodynamic limit.%
}. For large systems, both $\Omega_{max}$ and $\langle\Omega\rangle$
will equal $\omega_{0}\approx0.3868$ at $T=0$ and approach $\frac{6}{13}\sim0.46$
as $T\rightarrow\infty$. The first order nature of the transitions
between the different commensurate phases can be verified by looking
at the distribution of $\Omega$ near the transitions, which exhibits
two well separated peaks, with one overtaking the other as the transition
temperature is crossed \cite{futurepub}. This suggests that $\Omega_{max}(T)$
is a reliable indicator of these transitions. 

In Figure \ref{omega_max} we show curves of $\Omega_{max}$ and the
specific heat $C_{V}$\emph{ vs. T}, for a $150\times150$ system.
At low temperature, the system is in its ground state, and $\Omega_{max}=\omega_{0}$
(to within $\frac{1}{L}$). As temperature is increased, $\Omega_{max}$
undergoes three jump discontinuities before becoming fully continuous
as the rows are no longer homogeneous. While the step-wise behavior
of $\Omega_{max}$ is a clear indication of the existence of the commensurate
phases and the first order nature of the transitions between them,
its value at the plateaus does not give the exact periodicities due
to limited resolution and the existence of defects, and these then
must be inferred by looking at the configurations themselves, as in
Figure \ref{period8}. The first order transitions are accompanied
by small bumps in the specific heat at the same temperatures. In Figure
\ref{omega_max}, we identify phases with periods of $8$ and $13$
rows with the lower and middle plateaus, respectively. In our examinations
of systems of sizes up to $L=300$, we have identified phases with
$\Omega=\frac{1}{3},\frac{3}{8},\frac{5}{13},$ and $\frac{12}{31}$,
depending on system size and values of the coupling constants. At
these temperatures, we have verified numerically that the rows substantially
maintain their pure \emph{A} or \emph{B} character. The broad peak
in the specific heat at $T\sim.37$ attends the loss of purity in
the rows. 

These results may be understood by an effective coarse-grained description
of this system, appropriate for low temperatures, when the system
may be considered as composed of \emph{A }and \emph{B }type rows%
\footnote{Although this is easiest to justify when $J_{x}>J_{y}$, our numerical
study indicates that it has a large range of validity even when $J_{x}=J_{y}$.%
}. This description resembles the Frenkel-Kontorova model\cite{chaikin_principles_2000},
and exhibits a competition between length scales. To see this, note
that for the perfect\emph{ }tiling, the value of the frequency $\alpha_{n}$
for the $n^{th}$ row may be calculated from the tile markings: $\alpha_{n}=\lim_{L\rightarrow\infty}\frac{1}{L}\sum_{m=1}^{L}t_{mn}$.
From this, using Equations \ref{eq:Phi} and \ref{eq:theta}, we can
compute $\phi_{n}$. Now although defects enter the rows at finite
$T$, we may still define two frequencies, using the markings of the
top and bottom of a row. The ``top frequency'' is defined as $\alpha_{n}^{T}=\frac{1}{L}\sum_{m=1}^{L}t_{mn}$
while the ``bottom frequency'' is given by $\alpha_{n}^{B}=\frac{1}{L}\sum_{m=1}^{L}b_{mn}$.
For a perfect tiling, $\alpha_{n}^{B}=\alpha_{n+1}^{T}$, and therefore,
from Equation \ref{eq:mapping}, $\alpha_{n}^{B}/\alpha_{n}^{T}=q_{n}$,
but this will typically not be the case for defected tilings. This
notwithstanding, the $\phi_{n}$ at finite \emph{$T$ }may be inferred
from $\alpha_{n}^{T}$ using Equation \ref{eq:theta} in the same
manner as for the perfect tiling. These will be the variables used
in our effective description.

To construct an effective free energy for this model we assume that
the dominant contributions come from the entropy of the rows, and
the energy due to mismatches between the rows, each of which is characterized
by its frequencies $\alpha^{T}$ and $\alpha^{B}$ . We argue that
the energy cost associated with an imperfect interface between rows
$n$ and $n+1$ goes as $LJ_{y}\left|\alpha_{n}^{B}-\alpha_{n+1}^{T}\right|$,
where \emph{L} is the length of the row. Clearly, this term is zero
in the ground state, and numerical simulations bear out this functional
form for low temperatures\cite{futurepub}. At this level of coarse
graining, a row is characterized by its frequency $\alpha^{T}$ (or
equivalently $\alpha^{B}$), and its entropy should be a function
of this frequency which is extensive in \emph{L,} so that we shall
write the entropy of the $n^{th}$ row as $L\,\tilde{s}\left(\alpha_{n}^{T}\right)$. 

Taken together, we get that the free energy is given by $F/L=\sum_{n=1}^{L}J_{y}\left|q_{n}\alpha_{n}^{T}-\alpha_{n+1}^{T}\right|-T\tilde{s}\left(\alpha_{n}^{T}\right)$.
It is convenient to express this in terms of the variables $\phi_{n}$
discussed above. The free energy is then of the form

\begin{equation}
\frac{F}{L}=\sum_{m}J_{y}g\left(\phi_{m},\phi_{m+1}\right)-Ts\left(\phi_{m}\right)\label{eq:FreeEnergy-2}
\end{equation}
where $g\left(\phi_{m},\phi_{m+1}\right)$ is a function that favors
$\phi_{m+1}=\phi_{m}+\omega_{0}$, which holds identically in the
ground state. The entropy $s\left(\phi_{n}\right)$ depends only on
the fractional part $\left\{ \phi_{n}\right\} $, and thus it is a
periodic function with period one. It is the competition between these
two length scales which gives the novel behavior observed. Although
it is tempting to expand $g\left(\phi_{n},\phi_{n+1}\right)$ to first
order as $g\left(\phi_{n},\phi_{n+1}\right)\propto\left|\phi_{n+1}-\phi_{n}-\omega_{0}\right|$,
we note that such an expression fails when both $\left\{ \phi_{n}\right\} $
and $\left\{ \phi_{n+1}\right\} $ are larger than $1-\omega_{0}$
, since this would imply two adjacent \emph{B} rows, which carries
a disproportionately large energy cost.

The equilibrium configuration of the $\phi_{n}$ can be obtained by
minimizing $F$. As in the FK model, the first term favors an incommensurate
phase with a winding number $\omega_{0}$ while the entropy favors
a commensurate configuration. The temperature $T$ plays the role
of the strength of the periodic potential, so that commensurate phases
are expected at high temperature while incommensurate phases are expected
at low temperatures. The commensurate phases that are expected are
those with a winding number close to $\omega_{0}$ such as $1/3,\,3/8,\,5/13$,
\emph{etc}. We have observed some of these phases in our numerical
study, as seen in Figure \ref{omega_max}, where their presence is
indicated by the plateau values of $\Omega_{max}$. At still higher
temperatures the segregation into type \emph{A} and \emph{B} rows
breaks down, resulting in a disordered phase. It is interesting to
speculate about the low \emph{T }behavior of this system. It might
be that only at $T=0$ an incommensurate phase appears, but it could
be that such a phase, possibly with power-law correlations, is stable
at finite \emph{T. }These issues will be addressed in future work.

We would like to thank P. Chaikin, J. Kurchan, T. Lubensky, F. Sausset,
and G. Wolff for fruitful discussions. D.L. gratefully acknowledges
support from Israel Science Foundation grant 1574/08 and US-Israel
Binational Science Foundation grant 2008483.


\begin{thebibliography}{10}
\bibitem{fn1}
As distinct, for example, from the Heisenberg or x-y models, which employ
  continuous spins.
\bibitem{fn2}
Wang tilings employ square tiles, and as such may be used to model systems on a
  square lattice.
\end{thebibliography}
\end{document}